\documentclass[a4paper,english,aps,floats,twocolumn,showpacs,nofootinbib]{revtex4}
\usepackage{pslatex}
\usepackage[T1]{fontenc}
\usepackage{graphicx}
\usepackage{epsfig}

\usepackage{calc}
\usepackage{ifthen}
\usepackage{subfigure}

\newcommand{\bea}{\begin{eqnarray}}
\newcommand{\eea}{\end{eqnarray}}

\newcommand{\nc}{\newcommand}
\nc{\renc}{\renewcommand}
\nc{\eqs}[2]{\mbox{Eqs.~(\ref{#1},\,\ref{#2})}}
\nc{\eq}[1]{\mbox{Eq.~(\ref{#1})}}
\nc{\figs}[2]{\mbox{Figs.~(\ref{#1},\,\ref{#2})}}
\nc{\fig}[1]{\mbox{Fig~.(\ref{#1})}}
\nc{\be}[1]{\begin{equation} \mbox{$\label{#1}$}}
\nc{\ee}{\vspace{0.1cm}\end{equation}}

\newcommand{\bean}{\begin{eqnarray*}}
\newcommand{\eean}{\end{eqnarray*}}

\def\GeV{{\rm \ GeV}}

 \def\gae{\; ^{>}_{\sim} \;}

\nc{\npp}[3]{{\it  Nucl.\ Phys.\ }{{\bf #1} {(#2)} {#3}}}
\nc{\prdd}[3]{{\it  Phys.\ Rev.\ D\ }{{\bf #1} {(#2)} {#3}}}
\nc{\prll}[3]{{\it Phys.\ Rev.\ Lett.\ }{{\bf #1} {(#2)} {#3}}}
\nc{\pll}[3]{{\it  Phys.\ Lett.\ }{{\bf #1} {(#2)} {#3}}}

\begin{document}
\title{Comment on "Gravitino Dark Matter and Baryon Asymmetry from Q-ball Decay in Gauge Mediation (arXiv:1107.0403)"
}
\author{Francesca Doddato}
\email{f.doddato@lancaster.ac.uk}
\author{John McDonald}
\email{j.mcdonald@lancaster.ac.uk}
\affiliation{Cosmology and Astroparticle Physics Group, University of
Lancaster, Lancaster LA1 4YB, UK}
\begin{abstract}

    We comment on a statement made in arXiv:1107.0403, which implies that the model of gravitino dark matter from Q-ball decay we presented 
in arXiv:1101.5328 is generally ruled out by the effect of NLSP decay on BBN. We explain why this statement is incorrect.

 \end{abstract}
\maketitle

   In \cite{r1} we presented a model of late-decaying Q-balls as the source of gravitino dark matter in SUSY models with gauge-mediated SUSY breaking. This model was based on a large gravitino mass $m_{3/2} \approx 2 \GeV$. In this case the messenger scale is necessarily large, in which case the fragmentation of the Affleck-Dine condensate occurs for field values less than or of the order of the messenger scale, leading to unstable Q-balls.  Recently, an alternative model of gravitino dark matter from Q-ball decay \cite{s1} was studied in \cite{r2}, based on Q-balls which are unstable due to their small charge, rather than due to the large messenger mass scale as in our model. However, 
\cite{r2} contains a comment which effectively dismisses our model on the basis of NLSP decays and their effect on BBN. We do not agree with this comment.

             In \cite{r2}, it was stated that "The NLSP decays into the gravitino in the end, but they overlooked the fact that the NLSP abundance is limited severely by the BBN constriants for $m_{3/2} \gae 1 \GeV$, so that the amount of the produced gravitino should be much smaller than that to be the dark matter.". This implies that all NLSPs leading to a gravitino dark matter density with $m_{3/2} \gae 1 \GeV$ are excluded by BBN. The implication of this statement is that the late Q-ball decay mechanism for gravitino dark matter we presented in \cite{r1} is generally ruled out by BBN. We disagree with both the statement that $m_{3/2} \gae 1 \GeV$ is generally excluded by BBN constraints on NLSP decay and with the implication that our model is ruled out.

             The correct statement is that an MSSM NLSP with strict R-parity conservation and $m_{3/2} \gae 1 \GeV$ is excluded by BBN constraints \cite{r3}. 
However, this does not exclude all NLSP candidates:
\newline (1) Most notably, it does not exclude a RH sneutrino NLSP. In \cite{r4} the case of purely Dirac neutrinos was considered, in which case the MSSM LSPs (LSPs of the MSSM sector) which decay to RH sneutrino NLSPs are long-lived and BBN could be modified. However, it was found that BBN constraints are relatively weak and consistent with O(GeV) gravitino dark matter. We note that an even safer alternative would be to consider a weak-scale SUSY mass for RH sneutrinos, in which case the MSSM LSPs will decay to RH sneutrinos before nucleosynthesis as a result of the relatively large neutrino Yukawa coupling \cite{r5}.

\noindent (2) It does not exclude a conventional MSSM NLSP when there is a small amount of R-parity violation in the model \cite{r6}. For example, in the case of R-parity violating operators formed from Planck-scale suppressed operators and GUT-scale spontaneous R-parity violation, the MSSM NLSPs can decay sufficiently early to evade BBN constraints, but the gravitino remains sufficiently stable to serve as dark matter \cite{r6}. Again, our model would be perfectly viable in this case. 

We have taken the unusual decision to urgently clarify this issue by a brief comment, in order to counter the factually incorrect statement presented in \cite{r2} and its negative implications for our work. We will discuss these issues in greater detail in future work.


\end{document}